\documentclass{IEEEtran}
\usepackage{cite}
\usepackage{amsmath,amssymb,amsfonts}
\usepackage{algorithmic}
\usepackage{graphicx}
\usepackage{textcomp}
\usepackage[dvips]{epsfig}
\usepackage{color}
\newtheorem{theorem}{Theorem}

\newtheorem{remark}{Remark}

\newcommand{\proof}{\noindent \textit{Proof}: }
\newcommand{\carrew} {\hfill $\Box$}
\newcommand{\jgrv}[1]{{\color{red} #1}}

\newcommand{\carre}{\begin{flushright} \rule{2mm}{2mm} \end{flushright}}
\def\BibTeX{{\rm B\kern-.05em{\sc i\kern-.025em b}\kern-.08em
    T\kern-.1667em\lower.7ex\hbox{E}\kern-.125emX}}
\begin{document}
\title{Reformulation of Matching Equation in\\ Potential Energy Shaping}
\author{
    M.~Reza~J.~Harandi, and Hamid D. Taghirad \IEEEmembership{Senior Member, IEEE}
    \thanks{M.~Reza~J.~Harandi and Hamid D. Taghirad are with \textbf{A}dvanced
        \textbf{R}obotics and \textbf{A}utomated \textbf{S}ystems (ARAS),
        Faculty of Electrical Engineering,
        K. N. Toosi University of Technology,
        P.O. Box 16315-1355, Tehran, Iran.
        (email: jafari@email.kntu.ac.ir, taghirad@kntu.ac.ir)}
}

\maketitle
\begin{abstract}
Stabilization of an underactuated mechanical system may be
accomplished by energy shaping. Interconnection and damping
assignment passivity-based control is an approach based on total
energy shaping by assigning desired kinetic and potential energy to the
system. This method requires solving a partial differential equation
(PDE) related to  he potential energy shaping of the system. In this
short paper, we focus on the reformulation of this PDE to be solved
easier. For this purpose, under a certain condition that depends on
the physical parameters and the controller gains, it is
possible to merely solve the homogeneous part of potential energy
PDE. Furthermore, it is shown that the condition may be reduced into
a linear matrix inequality form. The results are applied to a number
of benchmark systems.
\end{abstract}

\begin{IEEEkeywords}
Total energy shaping, potential energy PDE, mechanical systems,
passivity-based control.
\end{IEEEkeywords}

\section{Introduction}
Stabilization of an underactuated mechanical system is a challenging
problem since it is not fully feedback linearizable due to fewer
number of actuators than the degrees of freedom (DOF). Nevertheless,
Interconnection and damping assignment passivity-based control
(IDA-PBC) is a general approach to stabilize an underactuated system
by shaping total energy of the system~\cite{ortega2002stabilization}.
In this method, the desired structure of the closed-loop system is
selected instead of classical passivity, and then all assignable
energy functions are derived. For this purpose, two sets of PDEs
related to the shaping of potential and kinetic energy functions,
which are called \emph{matching equations}, should be solved
analytically~\cite{ferguson2019matched}. By this means, the desired
inertia matrix should be derived from the kinetic energy PDE, and
then by knowing this matrix, the potential energy PDE is solved.
Since the analytic solutions of these equations shall be derived,
the applicability of this approach is significantly restricted.

Several papers have focused on solving or reformulating the matching
equations. In \cite{acosta2005interconnection}, solving the matching
equations of systems with one degree of underactuation has been
reported. It has been shown that upon certain conditions, the
solution of the kinetic and potential energy PDEs could be derived
analytically. Furthermore, Solving the PDEs of 2-DOF underactuated
systems satisfying some conditions has been reported
in~\cite{d2006further}. Simplification of kinetic energy PDE by
means of coordinate transformation has been studied
in~\cite{viola2007total}. In~\cite{donaire2015shaping} the total
energy shaping without solving the matching equations for a class of
underactuated systems has been investigated, (see also
\cite{romero2016energy}). A method for simplifying the kinetic
energy PDE has been reported in~\cite{harandi2021matching} in which
a particular structure for desired inertia matrix has been
considered to simplify the related matching equation. Recently in
\cite{harandi2021solution}, it has been shown that replacing a PDE with
some Pfaffian differential equations may ease up this stumbling
block of solving matching equations. Furthermore, it has
been shown that the solution of potential energy PDE may be divided
into homogeneous and non-homogeneous components,
while~\cite{j2021bounded} reports on the general structure of the
homogeneous solution. In some cases, derivation of desired potential
energy, especially the non-homogeneous component, is a prohibitive
task. The situation is more crucial when the kinetic energy PDE is
as challenging as to derive another desired inertia matrix for the
system.

In this note, we concentrate on reformulation of the potential
energy PDE. For this purpose, we derive a certain and applicable
condition for many systems to merely derive the homogeneous solution
of the PDE. Furthermore, we show that verification of the proposed
condition is quite easy since it is based on physical parameters of
the system, desired inertia matrix, and the controller gains
in the homogeneous part of desired potential function. Furthermore,
it is shown that the proposed condition may be reformulated into a
linear matrix inequality (LMI), that may impose some constraints on
the controller gains. Note that in this formulation, the gravity
torques of unactuated configuration variables remain in the
closed-loop system. Thus, the contribution of this note may be
interpreted as rejection of a specific unmatched disturbance by
merely suitable selection of the controller gains (see
\cite{Donaire-2019} for the challenges of unmatched disturbance
rejection). The effectiveness of the method is verified through some
examples.

\section{Review of IDA-PBC for Mechanical Systems}\label{s2}
Dynamic equations of a mechanical  system in port Hamiltonian
modeling are as follows~\cite{ortega2002stabilization}
    \begin{equation}
\label{1}
\begin{bmatrix}
\dot{q} \\ \dot{p}
\end{bmatrix}
=\begin{bmatrix}
0_{n\times n} & I_n \\ -I_n & 0_{n\times n}
\end{bmatrix}
\begin{bmatrix}
\nabla_q H \\ \nabla_p H
\end{bmatrix}
+\begin{bmatrix}
0_{n\times m} \\ G(q)
\end{bmatrix}
u,
\end{equation}
where $q,p\in\mathbb{R}^{n}$ are  generalized position and momentum,
respectively, $H(q,p)=\frac{1}{2}p^TM^{-1}(q)p+V(q)$ denotes the
Hamiltonian of the system which is the summation of kinetic and
potential energy, $0<M\in\mathbb{R}^{n\times n}$ denotes the inertia
matrix, $u\in\mathbb{R}^m$ is the input, $G(q)\in\mathbb{R}^{n\times
m}$ denotes full rank input mapping matrix and the operator $\nabla$
denotes gradient of a function which is represented by a column
vector. Presume that the target dynamic of the closed-loop system is
in the following form
    \begin{align}
\label{2}
\begin{bmatrix}
\dot{q} \\ \dot{p}
\end{bmatrix}
=
\begin{bmatrix}
0_{n\times n} & M^{-1}M_d \\ -M_dM^{-1} & J_2-GK_vG^T
\end{bmatrix}
\begin{bmatrix}
\nabla_q H_d \\ \nabla_p H_d
\end{bmatrix},
\end{align}
in which
$$H_d=\frac{1}{2}p^TM_d^{-1}(q)p+V_d(q),$$
is the summation of the desired kinetic and potential energy,
$J_2\in\mathbb{R}^{n\times n}$ is a free skew-symmetric matrix,
$K_v\in\mathbb{R}^{m\times m}$ is positive definite damping gain,
and $V_d$ should be designed such that $q_d=\text{arg   min} V_d(q)$
with $q_d$ being the desired equilibrium point. By setting (\ref{1})
equal to (\ref{2}), the control law is given as
\begin{align}
u&=(G^TG)^{-1}G^T(\nabla_q H-M_dM^{-1}\nabla_q H_d+J_2\nabla_p H_d)\nonumber\\&-K_vG^T\nabla_p H_d,\label{3}
\end{align}
while the following PDEs shall be satisfied
    \begin{align}
&G^\bot \{\nabla_q  \big(p^TM^{-1}(q)p\big) - M_dM^{-1}(q)\nabla_q \big(p^TM_d^{-1}(q)p\big)   \nonumber \\ &\hspace{2.9cm}+2J_2 M_d^{-1}p \} =0,   \label{4} \\
&G^\bot\{\nabla_q V(q)-M_dM^{-1}\nabla_q V_d(q)\} =0. \label{5}
\end{align}
The first PDE is related to kinetic energy shaping, and the second
one corresponds to the potential energy PDE. First, $M_d$ should be
derived from nonlinear PDE (\ref{4}) and then $V_d$ is obtained from
(\ref{5}). Stability of $q_d$ is ensured by considering $H_d$ as a
Lyapunov candidate whose derivative is $\dot{H}_d =-(\nabla_p H_d)^T
GK_vG^T\nabla_p H_d$. In the next
section, we focus on the reformulation of PDE (\ref{5}).

\section{Main Results}
As indicated in \cite{harandi2021solution} and~\cite{j2021bounded},
the solution of potential energy PDE can be divided to homogeneous
($V_{dh}$) and non-homogeneous ($V_{dn}$) components. Furthermore,
$V_{dh}$ is derived from the following PDE
\begin{align}\label{6}
G^\bot M_dM^{-1}\nabla_q V_d(q) =0,
\end{align}
and has the following form
$$V_{dh}=\phi(V_{dh_1},V_{dh_2},\dots),$$
in which, $V_{dh_i}$s are functions satisfying (\ref{6}). Typically,
the free design function $\phi$ is set as:
\begin{align}\label{7}
V_{dh}=\displaystyle\sum \frac{k_i}{2} (V_{dh_{i}}-V_{dh_{i}}^*)^2,
\end{align}
in which $V_{dh_{i}}^*=\left.V_{dh_{i}}\right|_{q=q_d}$ and $k_i$s
are free positive gains. Furthermore, $V_{dn}$ is the particular
solution of the PDE (\ref{5}).

The solution of (\ref{5}) clearly depends on $M_d$, which is derived
from PDE (\ref{4}). Since in general, the nonlinear PDE (\ref{4}) is
very hard to be solved analytically, especially in the cases that
the first term of (\ref{4}) is non-zero, deriving multiple $M_d$ is
a prohibitive task. This means that the designers are very reluctant
to solve (\ref{5}), given $M_d$. Furthermore, derivation of
non-homogeneous solution of (\ref{5}) is more challenging than that
of the homogeneous part. As an example, examine the case reported in
\cite{harandi2021matching} for the systems with $G=P[I_m,0_{m\times
n}]^T$ where $P$ denotes permutation matrix. In this case, the term
$G^\bot M_d$ in (\ref{6}) is a constant vector resulting in
straightforward PDE (\ref{6}) to be solved\footnote{Note that based
on the structure of $M_d^{-1}$ given in \cite{harandi2021matching},
the $k$th row of adjugate of $M_d^{-1}$ is constant, but its
determinant is configuration-dependent. However, the determinant is
omitted in (\ref{6}).}. However, for the same case derivation of the
non-homogeneous solution is still very difficult (see section
\ref{pen} for more details).

To tackle this problem, in the following theorem,  a condition is
proposed, through which it is enough to solve the PDE (\ref{6})
instead of (\ref{5}), and therefore, derivation of the homogeneous
solution would be sufficient.
\begin{theorem}\label{th1}
Consider IDA-PBC methodology introduced in section~\ref{s2} with
$V_d=V_{dh}$ as the solution of (\ref{6}). Then $q_d$ is stable if
the following condition is satisfied
    \begin{align}\label{8}
    \left.\Big(\frac{\partial^2 (V_{dh}+\eta)}{\partial q^2}\Big)\right|_{q=q_d}>0,
    \end{align}
in which,
    \begin{align}
    \eta:&=\int\displaystyle\sum_{i=1}^{n-m}\bigg(0.5\frac{( G_i^\bot\nabla V)  G_i^{\bot}}{\| G_i^{\bot}\|^2}\bigg){M_d}^{-1} M\dot{q}+0.5\dot{q}^TMM_d^{-1}\nonumber\\&\bigg(\frac{( G_i^\bot\nabla V)  G_i^{\bot^T}}{\| G_i^{\bot}\|^2}\bigg)dt,\label{9}
\end{align}
while $G_i^\bot$ denotes the $i$th row of $G^\bot$.\carre
\end{theorem}
Notice that verification of condition (\ref{8}) is quite simple
since only positive definiteness of a constant matrix
should be checked.

\proof Substitute the control law (\ref{3}) with $V_d=V_{dh}$ as the
solution of (\ref{6}) in (\ref{1}). Then, the closed-loop equations
are given by:
    \begin{align}
\begin{bmatrix}
\dot{q} \\ \dot{p}
\end{bmatrix}
&=
\begin{bmatrix}
0_{n\times n} & M^{-1}M_d \\ -M_dM^{-1} & J_2-GK_vG^T
\end{bmatrix}
\begin{bmatrix}
\nabla_q H_d \\ \nabla_p H_d
\end{bmatrix}\nonumber\\&-\begin{bmatrix}
 0 \\ \displaystyle\sum_{i=1}^{n-m}\frac{( G_i^\bot\nabla V)  G_i^{\bot ^T}}{\| G_i^{\bot }\|^2}
\end{bmatrix},
\label{10}
\end{align}
with $H_d=\frac{1}{2}p^TM_d^{-1}p+V_{dh}$. The last term in
(\ref{10}) is resulted from dissatisfaction of (\ref{5}) and is
interpreted as the natural gravity torques/forces of the system in
the direction of unactuated coordinates. Now, consider the following
function
\begin{align}\label{11}
\boldsymbol{V}=H_d+\eta.
\end{align}
It is easy to verify that its derivative along the trajectories of the system is
$$\dot{\boldsymbol{V}}=-(\nabla_p H_d)^T GK_vG^T\nabla_p H_d.$$
Thus, if it is  shown that $\boldsymbol{V}$ is a suitable Lyapunov
function, the proof is completed. For this purpose, we should show
the following terms
\begin{align}
\left.\frac{\partial \boldsymbol{V}}{\partial x}\right|_{x=x_d}=0, \quad \left.\frac{\partial^2 \boldsymbol{V}}{\partial x^2}\right|_{x=x_d}>0 \quad \mbox{with} \hspace{1mm}x=[q^T,p^T]^T.\label{12}
\end{align}
Note that $\eta$ may be represented by:
\begin{align*}
\eta&=\int_{q_0}^{q}\displaystyle\sum_{i=1}^{n-m}\bigg(\frac{( G_i^\bot(\nu)\nabla_\nu V(\nu))  G_i^{\bot}(\nu)}{\| G_i^{\bot}(\nu)\|^2}\bigg){M_d}^{-1}(\nu) M(\nu)d\nu.
\end{align*}
Hence, the only $p$-dependent term in $\boldsymbol{V}$ is
$\frac{1}{2}p^TM_d^{-1}p$ that clearly satisfies (\ref{12}).
Therefore, we verify conditions (\ref{12}) with respect to $q$ for
the other terms of $\boldsymbol{V}$. By this means, we have
\begin{align*}
&\left.\frac{\partial (V_{dh}+\eta)}{\partial q}\right|_{q=q_d}=\left.(\nabla V_{dh})^T\right|_{q=q_d}+\\& \displaystyle\sum_{i=1}^{n-m}\bigg(\frac{\big( G_i^\bot(q_d)\left.(\nabla V)\right|_{q=q_d} \big) G_i^{\bot}(q_d)}{\| G_i^{\bot}(q_d)\|^2}\bigg){M_d}^{-1}(q_d) M(q_d)=0.
\end{align*}
Recall that $q_d$ is an equilibrium point, and therefore, it is on
the manifold $G^\bot \nabla V=0$. Hence, the second condition of
(\ref{12}) will be reformulated as (\ref{8}) and this completes the
proof.\carrew

Notice that the closed-loop equations (\ref{10}) may be interpreted
as a simple IDA-PBC in the presence of unmatched disturbance. Thus,
it may be seen  as the rejection of particular form of external disturbance.
\begin{remark}
In the proof of Theorem~\ref{th1}, based on the IDA-PBC methodology,
it is shown that the closed-loop system is represented by
(\ref{10}).  Here, it is explained from another point of view.
 Suppose that (\ref{10}) is the target dynamic of the
closed-loop system. By multiplying (\ref{1}) and (\ref{10}) from
left side to the full rank matrix $[G,G^{\bot^T}]^T$, the control
law and kinetic energy PDE are derived as (\ref{3}) and (\ref{4}),
respectively, while $V_d$ is the solution of the PDE (\ref{6}).
\end{remark}

From (\ref{7}), it is deduced that the condition (\ref{8}) depends
on $M_d$ and $V_{dh}$, and may impose a constraint on $k_i$s.
Generally, the first term of (\ref{8}) is positive semi-definite;
thus, $M_d$ should be designed such that the second term is not
negative definite. In the following, some cases are analyzed.

Case 1: Presume that $q_d$ is natural equilibrium point of the
system and $M_d=M$. In this case, the condition (\ref{8}) is
trivially satisfied since it is the summation of two positive (semi)
definite matrices. The advantage of Theorem~\ref{th1} in this case
is the simple design of $V_d$ for the systems with $\left. \nabla
V\right|_{q=q_d}\neq 0$. This might be very useful in applications
such as underactuated cable-driven robots~\cite{harandi}.

Case 2: If $V_{dh}$ is designed as the general form of (\ref{7}),
the first term of (\ref{8}) is given by:
\begin{align*}
\left.\nabla^2V_{dh}\right|_{q=q_d}=\displaystyle\sum k_i\left.\begin{bmatrix}
\frac{\partial^2V_{dh_i}}{\partial q_1^2} & \dots & \frac{\partial(\partial V_{dh_i})}{\partial q_n\partial q_1} \\ \vdots & \ddots & \vdots \\ \frac{\partial(\partial V_{dh_i})}{\partial q_1\partial q_n} & \dots & \frac{\partial^2V_{dh_i}}{\partial q_n^2}
\end{bmatrix}\right|_{q=q_d}.
\end{align*}
Therefore, generally we can argue that the inequality (\ref{8}) can
be reformulated into an LMI. Solving a LMI problem has been studied
thoroughly in several references such as \cite{boyd1994linear}, and many
tractable software are developed to accomplish this task.

Case 3: Consider systems with $G=P[I_m,0_{m\times n}]^T$.  The
condition (\ref{8}) is in the following form
\begin{align*}
&\left.\nabla^2V_{dh}\right|_{q=q_d}+\displaystyle\sum_{i=1}^{n-m}\bigg(\left.\frac{\partial (G_i^\bot\nabla_{q}V)}{2\partial q}\right|_{q=q_d}G_i^\bot M_d^{-1}(q_d)M(q_d)\\&+M(q_d)M_d^{-1}(q_d)G_i^{\bot^T}\left.\Big(\frac{\partial (G_i^\bot\nabla_{q}V)}{2\partial q}\Big)^T\right|_{q=q_d}\bigg)>0
\end{align*}
In particular, assume that $n=2$ and $m=1$ which corresponds to most
of the benchmark systems given in the literature. In this case, we
have an inequality in the form of:
\begin{align}
\label{13}
k\begin{bmatrix}
\alpha_1 & \alpha_2 \\ \alpha_2 & a_3
\end{bmatrix}+\begin{bmatrix}
\beta_1 & \beta_2 \\ \beta_2 & \beta_3
\end{bmatrix}>0,
\end{align}
in which the first  and second matrix represent
$\left.\nabla^2V_{dh}\right|_{q=q_d}$ and
$\left.\frac{\partial^2\eta}{\partial q^2}\right|_{q=q_d}$,
respectively. The first matrix is assumed to be positive semi-definite with the following properties
\begin{align*}
\alpha_1,\alpha_3>0,\qquad \alpha_1\alpha_3-\alpha_2^2=0.
\end{align*}
As explained before, the second matrix must not be negative definite
(if so, $M_d$ should be altered). Therefore, we have the following
scenarios
\begin{align}
&A_1: \beta_1,\beta_3\leq0,\qquad \beta_1\beta_3-\beta_2^2\leq0\nonumber\\
&A_2: \beta_1,\beta_3\geq0,\qquad \beta_1\beta_3-\beta_2^2\leq0\nonumber\\
&A_3: \beta_1\beta_3\leq0.\label{14}
\end{align}
One can easily verify that (\ref{13}) is satisfied with
\begin{align}
&A_1\& A_3: k>\max\biggl\{-\frac{\beta_1}{\alpha_1},-\frac{\beta_3}{\alpha_3},\frac{\beta_2^2-\beta_1\beta_3}{\rho}\biggr\},\nonumber\\
&A_2:
k>\frac{\beta_2^2-\beta_1\beta_3}{\rho}, \label{15}
\end{align}
with 
$$\rho:=\alpha_1\beta_3+\alpha_3\beta_1-2\alpha_2\beta_2,$$ which must be positive. In the next
section, some examples are given as representative of benchmark
systems to show the applicability of the proposed method in
practice. Readers are referred to \jgrv{[arxiv version of this
note]} to see more examples.

\section{Case Studies}\label{s4}
\subsection{Spatial Underactuated Cable-Driven Robot}
This system consists of a suspended mass from two cables that their
length is controlled by the actuators. The dynamic parameters of the
system, as well as the solution of the potential energy PDE
(\ref{5}) with $M_d=M$ are given as  follows~\cite{harandi}
\begin{align*}
&q=[x,y,z]^T,\quad M=mI_3,\quad V=mgy,\quad q_d=[x_d,y_d,0]^T,\\ &l_1^2=x^2+y^2+z^2,\qquad l_2^2=(x-b)^2+y^2+z^2,\\
&G^T=\begin{bmatrix}
\frac{x}{l_1} & \frac{y}{l_1} & \frac{z}{l_1} \\
\frac{x-b}{l_2} & \frac{y}{l_2} & \frac{z}{l_2}
\end{bmatrix},\qquad V_d=mgy+\phi(x,y^2+z^2),
\end{align*}
in which $m$ denotes the mass of the end-effector while $y$ is
always negative. It is clear that by defining
$$V_{dh}=\frac{k_1}{2}(x-x_d)^2+\frac{k_2}{2}(y^2+z^2-y_d^2)^2,$$
it is not possible to satisfy $q_d=\text{arg   min} V_d(q)$.
However, by omitting $mgy$ and designing $V_d=V_{dh}$, the condition
(\ref{8}) is reduced to
 $$\begin{bmatrix}
 k_1 & 0 & 0 \\ 0 & 4k_2y_d^2 & 0 \\ 0 & 0 & 0
 \end{bmatrix}+\begin{bmatrix}
 0 & 0 & 0 \\ 0 & 0 & 0 \\ 0 & 0 & -\frac{mgy_d}{y_d^2+z_d^2}
 \end{bmatrix},$$
which shows that the condition (\ref{8}) of Theorem~\ref{th1} is satisfied with
$k_1,k_2>0$. This choice clearly simplifies the controller design.

\subsection{Acrobot}
The system is a 2R serial robot whose second link is merely
actuated. Dynamic parameters of the system are given
as~\cite{donaire2017robust}
\begin{align}
&M=\begin{bmatrix}
c_1+c_2+2c_3\cos(q_2) & c_2+c_3\cos(q_2) \\ c_2+c_3\cos(q_2) &  c_2
\end{bmatrix}\nonumber\\
&V=c_4g\cos(q_1)+c_5g\cos(q_1+q_2),\label{16}
\end{align}
and $G=[0,1]^T$. An IDA-PBC controller has been designed in
\cite{d2006further} with the following parameters for the controller
\begin{equation*}
\begin{array}{c}
M_d=\begin{bmatrix}
a_1 & a_2 \\ a_2 & a_3
\end{bmatrix},\\ V_d=b_0\cos(q_1-\mu q_2)+b_1\cos(q_1)+b_2\cos(q_1+q2)\\+b_3\cos(q_1+2q_2)+b_4\cos(q_1-q_2)+\phi(q_1-\mu q_2)
\end{array}
\end{equation*}
where
\begin{align*}
 a_3>\frac{a_2}{1-\sqrt{c_1/c_2}},\qquad
\mu=\frac{-1}{1+\sqrt{c1/c2}}.
\end{align*}
Furthermore, $b_0$ is a free constant and $b_i$s for
$i\in\{1,...,4\}$ are constant values defined in
\cite{donaire2017robust,d2006further}. To apply Theorem~\ref{th1},
consider the desired potential energy as follows
 $$V_d=V_{dh}=\frac{k}{2}(q_1-\mu q_2)^2.$$
Condition (\ref{8}) is in the form of (\ref{13}) with the following parameters
\begin{align*}
&\alpha_1=1,\qquad\alpha_2=-\mu,\qquad\alpha_3=\mu^2,\\& \beta_1=\frac{-c_4g-c_5g}{a_1a_3-a_2^2}(a_3c_1+a_3c_2+2a_3c_3-a_2c_2-a_2c_3),\\& \beta_2=\frac{-c_5g}{2(a_1a_3-a_2^2)}(a_3c_1+a_3c_2+2a_3c_3-a_2c_2-a_2c_3)\\&+\frac{-c_4g-c_5g}{2(a_1a_3-a_2^2)}(a_3c_2+a_3c_3-a_2c_2), \\&\beta_3=\frac{-c_5g}{a_1a_3-a_2^2}(a_3c_2+a_3c_3-a_2c_2)
\end{align*}
Since $a_3>a_2$, the case $A1$ in (\ref{14}) is applicable 
and the suitable value of $k$ is given in (\ref{15}). Note that
since $a_i$s are all positive free scalars, assurance of $\rho>0$ is
quite simple.

\subsection{Pendubot}\label{pen}
As explained before, Theorem~\ref{th1} is appropriate to integrate
with  the proposed method in~\cite{harandi2021matching} in which a
systematic approach to solve or simplify (\ref{4}) has been
introduced. The reason is the particular structure of $M_d$ such
that the vector $\gamma$ defined as
\begin{align*}
\gamma:=\det M \det M_d^{-1}G^\bot M_dM^{-1},
\end{align*}
is independent of the configuration-dependent element of $M_d$. By
this means, $M_d$ and $V_{dh}$ may be derived easily. However,
derivation of $V_{dn}$ is very complicated in general. As an
example, consider Pendubot whose dynamic parameters are given as
(\ref{16}) with $G=[1,0]^T$.
By applying the procedure of \cite{harandi2021matching}, $M_d^{-1}$ is derived as follows
\begin{align*}
M_d^{-1}=\begin{bmatrix}
a_1 & b_1 \\ b_1 & \frac{\lambda e^{a(q_2)}+b^2}{a_1}
\end{bmatrix}
\end{align*}
in which
\begin{align*}
&a(q_2)=\frac{-a_1c_1c_2c_3+b_1c_1c_2c_3}{c_1c_2(b_1c_3+2a_1c_3)}\ln\big(c_1c_2-c_3^2\cos^2(q_2)\big)+\\&\frac{2a_1c_1c_2c_3-2b_1c_1c_2c_3}{c_1c_2(b_1c_3+2a_1c_3)}\cos(q_2)+\ln\bigg(\frac{1+\sqrt{\frac{c_3^2}{c_1c_2}}\cos(q_2)}{1-\sqrt{\frac{c_3^2}{c_1c_2}}\cos(q_2)}\bigg)\times\\&\frac{2a_1c_3^3+2a_1c_3^2(c_1+c_2)+2b_1c_2c_3^2(b_1-2)}{2a_1\sqrt{c_1c_2c_3^2}(b_1c_3+2a_1c_3)}\\&+\frac{b_1c_3^3}{c_3^2(b_1c_3+2a_1c_3)}\ln\big(c_3^2\cos^2(q_2)-c_1c_2\big),
\end{align*}
$\lambda$ is a free parameter,
and the constants $a_1$ and $b$ should be chosen such that 
$a_1c_1+a_1c_2+b_1c_2=0$. The solution of PDE (\ref{6}) which is independent of $a(q_2)$ is
\begin{align*}
V_{dh}=\phi\bigg(\frac{\ln\big(\frac{\delta_4+\delta_3\cos(q_2)+\sqrt{\delta_4^2-\delta_3^2}\sin(q_2)}{(\delta_3+\delta_4\cos(q_2))(\delta_1\delta_4-\delta_2\delta_3)}\big)}{\delta_4\sqrt{\delta_4^2-\delta_3^2}}+\frac{\delta_2}{\delta_4}q_2-q_1\bigg)
\end{align*}
with
\begin{align*}
&\delta_1=-b_1c_2-a_1c_2,\qquad\hspace{.8cm}\delta_2=-a_1c_3,\\
&\delta_3=b_1c_2+a_1c_1+a_1c_2,\qquad\delta_4=b_1c_3+2a_1c_3.
\end{align*}
However, derivation of the non-homogeneous solution of (\ref{5}) is
a prohibitive task. To tackle this problem, we may investigate whether
Theorem~\ref{th1} could be applied in this case. For this purpose,
by virtue of (\ref{7}), condition (\ref{8}) is in the following form
\begin{align*}
&\alpha_1=1,\qquad\alpha_2=-\frac{1}{\delta_4(\delta_3+\delta_4)},\qquad \alpha_3=\alpha_2^2,\\
&\beta_1=-b_1c_5g(c_1+c_2+2c_3)-\frac{\lambda a(0)+b^2}{a_1}c_5g(c_2+c_3),\\
&\beta_2=-b_1c_5g(c_2+c_3)/2-\frac{\lambda a(0)+b^2}{2a_1}c_2c_5g-b_1c_5g(c_1+c_2\\&+2c_3)/2-\frac{\lambda a(0)+b^2}{2a_1}c_5g(c_2+c_3),\\
&\beta_3=-b_1c_5g(c_2+c_3)-\frac{\lambda a(0)+b^2}{a_1}c_2c_5g.
\end{align*}
It seems that condition (\ref{8}) coincides with $A1$ in (\ref{14}) that the suitable value of $k$ is derived from (\ref{15}). Note that $\lambda,a_1$ and $b$ should be chosen such that $\rho>0$. 

As a numerical example, similar to \cite{harandi2021matching},  presume that $c_1=4,c_2=1,c_3=1.5,c_5=2$. 
Condition (\ref{8}) is in the following form
\begin{align*}
k\begin{bmatrix}
1 & 5/9 \\ 5/9 & 25/81
\end{bmatrix}+\begin{bmatrix}
-550 & -420 \\ -420 & -290
\end{bmatrix}
\end{align*}
which is matched with $A1$ in (\ref{14}). The suitable value of
$k$ is derived form (\ref{15}) with $\rho=6.91$.
\subsection{Cart-pole}
The system consists of a suspended mass from a cart. Dynamic equations are as follows~\cite{acosta2005interconnection}
\begin{align*}
M&=\begin{bmatrix}
1 & b\cos(q_1) \\ b\cos(q_1) & m_3
\end{bmatrix},\quad V=c\cos(q_1),\quad G=[0,1]^T\\&
c=g/l,\quad b=1/l,m_3=(m+M)/ml^2,\quad q=[\theta,x]^T.
\end{align*}
In literature, the designed IDA-PBC are based on a partial feedback linearization to transform the system into so called Spong's normal form. Here, our goal is to stabilize the system without partial feedback linearization. For this purpose, we apply the method introduced in \cite{harandi2021matching}. By this means, $M_d^{-1}$ is in the following form 
\begin{align*}
M_d^{-1}=\begin{bmatrix}
\frac{\lambda e^{a(q_2)}+b^2}{a_1} & b_1 \\ b_1 & a_1
\end{bmatrix}
\end{align*}
 with
\begin{align*}
&a(q_1)=\frac{2\ln\big(bb_1+a_1m_3(a_1m_3-bb_1)\tan^2(q_1/2)\big)}{a_1(m_3a_1^2-b_1^2)}\times\\&(m_3a_1^2-2m_3a_1b_1+b_1^2)-\\&\frac{ln\big(b\tan^2(q_1/2)-b+\sqrt{m}+\sqrt{m}\tan^2(x/2)\big)}{a_1^3m_3+a_1^2b_1\sqrt{m_3}}\times\\&(m_3a_1^2-2m_3a_1b_1+b_1^2)-\\&\frac{\ln\big(b-b\tan^2(x/2)+\sqrt{m}+\sqrt{m}\tan^2(x/2)\big)}{a_1^3m_3+a_1^2b_1\sqrt{m_3}}\times \\&(m_3a_1^2-2m_3a_1b_1+b_1^2)
\end{align*}
and $V_{dh}$ is
\begin{align*}
&V_{dh}=\phi\bigg(-\frac{\ln\Big(\frac{bb_1+\sin(q_1)\sqrt{-a_1^2m_3^2+b^2b_1^2}+a_1m_3\cos(q_1)}{a_1m_3+bb_1\cos(q_1)}\Big)}{bb_1\sqrt{-a_1^2m_3^2+b^2b_1^2}}\times\\&(-bm_3a_1^2+bb_1^2)-\frac{a_1q_1}{b_1}-q_2\bigg).
\end{align*}
Condition (\ref{8}) is in the form of (\ref{13}) with the following parameters
\begin{align*}
&\alpha_1=\alpha_2^2,\alpha_2=\frac{-bm_3a_1^2+bb_1^2}{bb_1(a_1m_3+bb_1)+\frac{a_1}{b_1}},\alpha_3=1,\\&
\beta_1=-c\frac{\lambda e^{a(0)}+b^2}{a_1}-bb_1c,\beta_2=\frac{\beta_1+\beta_3}{2},\\&\beta_3=-bc\frac{\lambda e^{a(0)}+b^2}{a_1}-b_1m_3c
\end{align*}
It is matched with $A1$ in (\ref{13}) and the suitable value of $k$ is derived from (\ref{15}) where positiveness of $\rho$ may be deduced from suitable values of $a_1,b_1$ and $\lambda$.
\subsection{VTOL Aircraft}
It is a strongly coupled system with the following modified dynamics~\cite{acosta2005interconnection}
\begin{align*}
M=I_3,V=\frac{g}{\epsilon}\cos(q_3),G=\begin{bmatrix}
1 & 0 \\ 0 & 1 \\ \frac{1}{\epsilon}\cos(q_3) & \frac{1}{\epsilon}\sin(q_3)
\end{bmatrix}q=\begin{bmatrix}
x \\ y \\ \theta
\end{bmatrix}
\end{align*}
Here our aim is to modify the proposed controller in \cite{acosta2005interconnection}. 
The parameters of IDA-PBC controller are
\begin{align*}
&M_d^{-1}=\begin{bmatrix}
\lambda_1\epsilon\cos^2(q_3)+\lambda_3 & \lambda_1\epsilon\cos(q_3)\sin(q_3) & \lambda_1\cos(q_3) \\ \lambda_1\epsilon\cos(q_3)\sin(q_3) & -\lambda_1\epsilon\cos^2(q_3)+\lambda_3 & \lambda_1\sin(q_3) \\ \lambda_1\cos(q_3) & \lambda_1\sin(q_3) & \lambda_2
\end{bmatrix}\\&
V_{dh}=\phi(q_1-q_{1d}-\frac{\lambda_3}{\lambda_1-\lambda_2\epsilon}\sin(q_3),q_2-q_{2d}+\\&\frac{\lambda_3-\lambda_1\epsilon}{\lambda_1-\lambda_2\epsilon}(\cos(q_3)-1))
\end{align*}
in which the following inequality should be held
$$\lambda_3>5\lambda_1\epsilon,\quad \lambda_1/\epsilon>\lambda_2>\lambda_1/2\epsilon.$$
Condition (\ref{8}) in this case is as follows
\begin{align*}
&k_1\begin{bmatrix}
1 & 0 & -\frac{\lambda_3}{\lambda_1-\lambda_2\epsilon} \\ 0 & 0 & 0 \\ -\frac{\lambda_3}{\lambda_1-\lambda_2\epsilon} & 0 & \big(\frac{\lambda_3}{\lambda_1-\lambda_2\epsilon}\big)^2
\end{bmatrix}+k_2\begin{bmatrix}
0 & 0 & 0 \\ 0 & 1 & 0 \\ 0 & 0 & 0
\end{bmatrix}+\\&\begin{bmatrix}
0 & 0 & 0\\ 0 & 0 & 0 \\ \theta_1 & 0 & \theta_2
\end{bmatrix}>0
\end{align*}
with
\begin{align*}
\theta_1&=\frac{g\epsilon(-\lambda_1\lambda_2\epsilon+\lambda_2\lambda_3-\lambda_1^2\epsilon^2+\lambda_1\lambda_3\epsilon)}{(\lambda_1\epsilon+\lambda_3)(-\lambda_1\epsilon+\lambda_3)\lambda_2-\lambda_1(-\lambda_1\epsilon+\lambda_3)\lambda_1}\\
\theta_2&=\frac{g\epsilon(\lambda_1^2\epsilon-\lambda_1\lambda_3+\lambda_1^2\epsilon^3-k_3^2\epsilon)}{(\lambda_1\epsilon+\lambda_3)(-\lambda_1\epsilon+\lambda_3)\lambda_2-\lambda_1(-\lambda_1\epsilon+\lambda_3)\lambda_1}
\end{align*}
that can be solved easily.
\section{Conclusion and Future Works}
In this paper, we concentrated on the reformulation of potential energy PDE of IDA-PBC approach. It was shown that under the satisfaction of a condition, it is enough to merely derive the homogeneous solution of this PDE. The condition was analyzed for three cases including 2-DOF underactuated robots. Furthermore, it was deduced that the proposed condition may be reformulated into LMI problem. The proposed method was verified on an underactuated cable-driven robot, Acrobot, and Pendubot. Generalization of the method to simplify the kinetic energy PDE is our future aim.   
\bibliographystyle{ieeetr}
\bibliography{ref}
\end{document}